\documentclass[a4paper,usenatbib]{mnras}

\usepackage{mathptmx}

\usepackage[T1]{fontenc}
\usepackage{ae,aecompl}

\usepackage{amsmath}	
\usepackage{amssymb}	
\usepackage{graphicx}	

\bfseries

\title[MOND and SIV theory]{MOND as a peculiar case of the SIV theory}
\author[Maeder]{Andre Maeder \thanks{E-mail: Andre.Maeder at UniGe.ch}\\
{Geneva Observatory - chemin Pegasi 51, CH-1290 Sauverny, Switzerland} }

\date{Accepted 2023 January 3. Received 2022 December 13; in original form 2022 October 20}

\pubyear{2021}

\begin{document}
\label{firstpage}
\pagerange{\pageref{firstpage}--\pageref{lastpage}}
\maketitle


\begin{abstract}
The scale invariant theory is preserving the fundamental physical properties of General Relativity, 
while enlarging the group of invariances subtending  gravitation theory  (Dirac1973; Canuto et al.1977).
The Scale Invariant Vacuum (SIV) theory  assumes, as gauging condition,  that 
``The macroscopic empty space is scale invariant, homogeneous and isotropic''. 
 Some basic properties in Weyl's Integrable Geometry and cotensor calculus  are examined
  in relation with  scalar-tensor theories.
Possible scale invariant effects are   strongly reduced by matter density, 
both at the cosmological  and local  levels.
The weak field limit of SIV  tends to MOND, when the scale factor is taken as constant, 
an approximation valid  (<1\%) over the last 400 Myr.
A better understanding  of the  $a_0$-parameter is obtained:  it corresponds
  to   the equilibrium point of   the   Newtonian and SIV dynamical  acceleration.
Parameter $a_0$ is not a universal constant, it depends on the density and  age of the Universe.
As MOND is doing, SIV theory avoids the call to dark matter, moreover  the cosmological models 
 predict  accelerated expansion.
\end{abstract}


\section{Introduction} \label{intro}

In the context of the dark matter problem, 
the  Modified Newtonian Dynamics (MOND)  was  proposed by \citep{Milgrom83}; this dynamics was  accounting  for the flat 
rotation curves of galaxies. 
Over the following decades, this theory received a number of further  extensions and applications,
{\it{e.g.}}\citet{Milgrom09,Milgrom14a,Milgrom14b}.  
The application of MOND to observations 
is meeting a number of positive results, see  review by \citet{FamaeyMc12}. 
Agreement has  been obtained for the flat rotation curves of spiral galaxies \citep{Lelli17},
 also  for clusters of galaxies  \citep{Sanders03,Milgrom18}, as well as
in the Local Group  \citep{Pawlowski12,PawlowskiK22} and  in the Fornax Cluster    \citep{Asencio22}.

 The dynamical acceleration  ($ V^2/R$, where $V$ is the velocity and 
$R$ the galactocentric distance)  is  related  to the baryonic 
 gravitiy ($GM/R^2$)  for spiral, irregular, elliptical, lenticular and spheroidals by a unique thin relation that significantly deviates from the
 Newtonian Law at very low gravities  \citep{McGaugh16,Lelli17,Li18}.
The unicity of this relation  is  striking, because it concerns observations made in different types of galaxies 
and  at different distances from their centers.
Such a result is uneasy to account for  by  dark matter, and it has been  recognized to better correspond to a gravity effect \citep{Lelli17}.


In MOND, the usual Newtonian gravity law remains unmodified for gravities above  a constant value 
 $a_0  \approx 1.2  \cdot 10^{-8} $ cm s$^{-2}$.
In the so-called  {\it{deep-MOND limit}}, for gravities  much lower than  $a_0$,
 a different gravitation law  applies with a gravity $g$ given by,
\begin{equation}
g \, = \, \sqrt{a_0 \; g_{\mathrm{N}}} \, ,
\label{deep}
\end{equation}
\noindent
where $g_{\mathrm{N}}$ is the usual Newtonian gravity. 

 In attempts  to  generalize the theory, developments of the basic MOND have been made in 
 a variety of theoretical  directions \citep{Milgrom15}:
  in the non-relativistic regime  the 
 {\it{Modified Poisson Gravity}} and {\it{ Quasilinear MOND}}, both  space dilation invariant; in the relativistc regime  the
 {\it{Tensor-Vector-Scalar (TeVeS), MOND adaptations of Einstein–Aether theories,  Bimetric MOND gravity (BIMOND) \citep{Milgrom22}}}.
 Properties of General Relativity (GR), such as the weak  Principle of Equivalence, the Lorentz invariance and 
  the general covariance, become
  invalid in some of the MOND extensions \citep{Milgrom19}.\\

Scale invariance was first considered by \citet{Weyl23} and \citet{Eddington23} in order to account 
by the geometry of space-time for both gravitation and electromagnetism, before being  abandonned because the properties of a particle 
would have depended on its past worldline  \citep{Einstein18}. A revival of the theory was first brought about by \citet{Dirac73}
and then by \citet{Canuto77}, who considered the so-called Weyl Integrable Geometry (WIG),  where Einstein's criticism
does not apply (see Sect. 2.2). 
\citet{Dirac73} emphasized that: {\it{``It appears as one of the fundamental principles
in Nature that the equations expressing basic laws should be invariant
under the widest possible group of transformations''}}. 
Scale invariance means that the basic equations do not change upon a transformation 
of the line element of the form, 
\begin{equation}
ds'\,=\,\lambda(x^{\mu})\,ds\, ,
\label{ds}
\end{equation}
where $\lambda(x^{\mu})$ is the scale factor,  $ds'$ refers to GR while $ds$ refers  to the WIG space, for example.
Scale invariance is present in Maxwell's equations 
in absence of charge and currents, as well as in General Relativity (GR) 
in the case of  the empty space \citep{Bondi90}.  

Geodesics and geometrical properties of the WIG were studied by \citet{BouvierM78} and   the weak field limit, which shows
an additional acceleration in the direction of motion, by  \citet{MBouvier79}.   
Instead of the Large Number Hypothesis \citep{Dirac74},  a new gauging condition is now
applied to derive  the cosmological equations, which naturally show an accelerated expansion;  
several observational tests were performed \citep{Maeder17a,Maeder17b}.  A number of studies followed: 
 on the rotation of galaxies and the dynamics of clusters of galaxies
\citep{Maeder17c,MaedGueor20b}; on the growth of density fluctuations in the early Universe, where the formation of galaxies 
is favoured by the additional acceleration without the need of dark matter \citep{MaedGueor19}; 
a general review \citep{MaedGueor20a}; on horizons and inflation \citep{MaedGueor21a} and  on the lunar recession \citep{MaedGueor22}.
The tests were positive and cast doubts on the need of dark components. 

The aim of this work is to study the relations between MOND and SIV theories. 
In Section 2, we recall the basics of the SIV theory  and examine 
the relation with scalar-tensor theories. 
  In Section 3,  some cosmological  properties of the  SIV theory are emphasized 
  in view of Section 4, where the Newtonian and MOND 
 approximations in the SIV theory are derived. The meaning and numerical value  of the $a_0$-parameter  are examined.
  Section 5 contains the conclusions.    
 
 
 \section{The basic theoretical context}
 
 \subsection{Scalar-tensor  versus  cotensor theories}   \label{Scotensor}

 Among alternative theories of gravity, the scalar-tensor theories represent 
 one of  the most elaborated developments\footnote{See a  
 312 pages review by Clifton, Ferreira, Padilla and Skordis on {\emph{``Modified Gravity and Cosmology''}} \citep{Clifton12}}.
 In particular, the scalar-tensor  theories also  offer an appropriate framework for the implementation of scale invariance.
 A fundamental property of these theories is the presence of an extra scalar field  $\varphi$, 
  acting in  parameter domains where General Relativity
 is not currently tested. The occurrence of dark components  
 has  stimulated such researches.
 The starting point is generally the expression   of the action, 
 
 \begin{equation}
 S\, = \, \int d^4x \sqrt{-g}\left[\frac{1}{12} \alpha \varphi^2 R+
  \frac{1}{2} \partial _{\mu}\varphi\partial^{\mu} \varphi+ \lambda \varphi^4 
 \right]\,,
 \label{actionF}
 \end{equation}
 
 \noindent
 here in a simple version given by  \citet{FerreiraT20}, with $\varphi$ the scalar field.
 As pointed out by these authors, the theory is invariant under $g_{\mu \nu}
 \rightarrow  \lambda^2   g_{\mu \nu}$ and $\varphi \rightarrow \lambda^{-1} \varphi$ 
 where $\lambda$ is a constant. With $\alpha=1$, the
 theory is conformally invariant.
   Decades ago, applications to varying $G$ were often considered (see also \citet{Fujii03}),  
   while present perspectives more concern  early phases and
    inflation as well as advanced  phases of accelerated expansion. Observational signatures of scale invariant gravity 
    predicted in the scalar-tensor theories  are quite rare,
     however a pertinent signature to distinguish such effects from those of General Relativity may 
   be provided  by perturbations in the ring down phase of black holes, with signatures in their resulting 
   gravitational wave emission  \citep{FerreiraT20}.\\
  
  In the scale invariant vacuum (SIV) theory, an action with great similarities  to the above 
  Eq.(\ref{actionF}) can also  be expressed, see below expression (\ref{act2}).
  However, rather than a scalar-tensor theory, SIV should better be called {\emph{``a cotensor theory''}}.
  Now, some preliminaries  are required. SIV  applies properties of  the Weyl's Geometry
   \citep{Weyl23,Eddington23,Dirac73}, where  the metrical determination is given
   by the usual quadratic form $ds^2= g_{\mu\nu} dx^{\mu}dx^{\nu}$. Moreover,  
    the length $\ell$ of any vector with contravariant components $a^{\mu}$ is determined by a scale factor $\lambda(x^{\mu})$
  and the same for the line element $ds$,
  \begin{equation}
  \ell^2 \, = \, \lambda^2(x^{\mu}) \, g_{\mu\nu} \, a^{\mu} a^{\nu}\,,  \quad \mathrm{and}
   \; \; ds'=\lambda (x^{\mu})  \, ds\,.
  \label{ell}
  \end{equation}
  This last relation with the quadratic form implies 
  that  Weyl's space is conformally equivalent to an other space\footnote{In the original Weyl's Geometry, this other space was not
  necessarily General Relativity, while in WIG it is always the case.} (defined by $g'_{\mu\nu}$)
through the  gauge transformation,
\begin{equation}
g'_{\mu\nu}\,=\,\lambda^{2}\,g_{\mu\nu}\,.
\label{conformal}
\end{equation}
 In the transport from a point 
$P_1(x^{\mu})$ to a nearby point  $P_2(x^{\mu}+dx^{\mu})$, the length
$\ell$ of a vector is assumed to change by,
\begin{equation}
d\ell \, = \, \ell \, \kappa_{\nu} \, dx^{\nu}\, .
\label{delu}
\end{equation}
There, $\kappa_{\nu}$ is called  the coefficient of metrical connection,  as a fundamental characteristics of the geometry
alike  $g_{\mu \nu}$. If we change the standards of length $\ell$  of any vector to $\ell'= \lambda(x)\,\ell$, one has
\begin{eqnarray}
\ell'+d\ell'= (\ell+d\ell) \, \lambda(x+dx) =(\ell+d\ell) \lambda(x) + \ell \, \lambda,_{\mu} dx^{\mu}\,,\\ \nonumber
\mathrm{thus} \; \; d\ell'=  \lambda(x)\, d\ell + \ell \, \lambda,_{\mu} dx^{\mu}= 
\lambda(x) \ell \, (\kappa_{\mu} + \Phi,_{\mu}) dx^{\mu}\,, \\ \nonumber
\mathrm{with} \; \;  \Phi,_{\mu}\,=\,  \frac{\lambda,_{\mu}}{\lambda}\,,\; \; \mathrm{and} \; \, \Phi \,= \,  \ln \lambda. 
\label{fnu}
\end{eqnarray}
Thus, we get for  $ d\ell'=\ell' \, \kappa'_{\nu} \, dx^{\mu}\,,$  (note that different derivatives, e.g. with respect to $x^{\mu}$ are indicated
 by ``$_{,\mu}$ ``,  or  a by  ``$_ {;\mu}$ `` or even  simply by ``$_{\mu}$'', see definitions by \citet{Dirac73}),
\begin{equation}
\kappa'_{\nu}= \kappa_{\nu} \, + \Phi,_{\mu}\,.
\label{knu}
\end{equation}
 
 A quantity $Y$, scalar, vector or tensor, which in a scale transformation changes  like $Y' \,= \, \lambda^n(x) \, Y$ is said to 
 be {\emph{coscalar, covector or cotensor }} of  power $\Pi(Y) =n$, this is called {\emph{scale covariance}}. For $n=0$,
  we have an {\emph{inscalar, invector or intensor}},  this is the particular case of scale invariance.
  The derivatives  $Y,_{\mu}$   do not necessarily enjoy
 the {\emph{co-covariant}}  (a defintion by Dirac) or invariant properties,
   but  derivatives  with such properties can be defined. Let us take for example a scalar $S$ of power $n$.
  Its ordinary (covariant) derivative is $S,_{\mu}$ (also noted $S_{\mu}$ when no ambiguity).
   Following \citet{Dirac73}, let us perform a change of  scale, we have
  \begin{equation}
  S'_{\mu}\,= \, (\lambda^n S)_{\mu}= \lambda^n S_{\mu}+ n \lambda^{n-1} \lambda_{\mu} S=
   \lambda^n (S_{\mu}+ n\, \Phi_{\mu}) S\,.
   \label{fnu}
   \end{equation}
   With Eqs. (\ref{knu}) and (\ref{fnu}), we get
   \begin{equation}
 (S_{\mu} - n \kappa _{\mu} S)' \, = \, \lambda^n (S_{\mu} - n \kappa _{\mu} S)   \, .
 \label{S}
 \end{equation}
 Now, we see that the covector  $S_{*\mu}=  (S_{\mu} - n \kappa _{\mu} S)$  of power $n$ 
 is the co-covariant derivative of scalar $S$  (such derivatives are always indicated with   ``$_*$'').
This ensures that the 
co-covariant derivatives preserve the power of the object they are applied on, thus the co-covariant derivative is also preserving 
scale covariance.

For a covector of power $n$, similar developments  \citep{Dirac73} lead to the derivatives of covectors,
\begin{eqnarray}
A_{\mu*\nu}\, & = & \,\partial_{\nu}A_{\mu}-^{*}\Gamma_{\mu\nu}^{\alpha}A_{\alpha}-n\kappa_{\nu}A_{\mu},\,\\{}
A_{*\nu}^{\mu}\, & = & \,\partial_{\nu}A^{\mu}+
^{*}\Gamma_{\nu\alpha}^{\mu}A^{\alpha}-n\kappa_{\nu}A^{\mu},\,\label{eq:co-cov_der}\\
\mathrm{with}\quad^{*}\Gamma_{\mu\nu}^{\alpha} & = & \Gamma_{\mu\nu}^{\alpha}+g_{\mu\nu}\kappa^{\alpha}\,-g_{\mu}^{\alpha}\kappa_{\nu}-g_{\nu}^{\alpha}\kappa_{\mu}, \label{eqco}
\end{eqnarray}
\vspace{-4mm}
\begin{equation}
\mathrm{with}  \quad   ^{*}\Gamma_{\sigma,\mu \nu}=  \, g _{\sigma \alpha}^{*}\Gamma_{\mu\nu}^{\alpha}\,.
\end{equation}
There,  $^{*}\Gamma_{\mu\nu}^{\alpha}$ is a modified Christoffel symbol,
while $\Gamma_{\mu\nu}^{\alpha}$ is the usual form. As a further example, 
the first derivatives of a co-tensor have the following expressions,
\begin{eqnarray}
T^{\mu \nu}_{* \rho} =T^{\mu \nu}_{, \rho}+^{*}\Gamma_{\rho\sigma}^{\mu} T^{\sigma \nu}+
^{*}\Gamma_{\rho\sigma}^{\nu} T^{\mu\sigma}-n \kappa_{\rho}T^{\mu \nu} \; ,\\
T_{\mu\nu * \rho} =T_{\mu\nu , \rho}- ^{*}\Gamma_{\mu \rho}^{\sigma} T_{\sigma \nu}-
^{*}\Gamma_{\nu\rho}^{\sigma} T_{\mu\sigma}-n \kappa_{\rho}T_{\mu \nu} \, ,
\end{eqnarray}
which also ensures their scale covariance.

Second co-covariant  derivatives can also be expressed. As an example, the second derivative
of coscalar $S$ becomes,

\begin{equation}
S_{*\mu * \nu} \, = \, S_{*\mu ,\nu} - (n-1) \kappa_{\nu} S_{*\mu} +\kappa_{\mu} S_{*\nu}- 
g_{\mu \nu} \kappa^{\sigma} S_{*\sigma} \, .
\end{equation}
\noindent
 An appropriate expression for the second derivative  of  a covector
can also be derived.  See  \citet{Canuto77} for a  summary of cotensor calculus.
 Operations on  covectors and cotensors can also be performed. Briefly, 
a contravariant covector $a^{\mu}$  with power $m$ multiplied
by a contravariant vector $b^{\nu}$ with power $n$ will form  a cotensor
$a^{\mu}b^{\nu} $ of  power (m+n). The corresponding covariant 
components $a_{\mu}$ and  $b_{\nu}$ have power (m+2) and (n+2) respectively. 
Also, the scalar
product $\vec{a} \cdot \vec{b}= g_{\mu \nu} a^{\mu} b^{\nu}$ gives a coscalar of power (m+n+2), the angle $\vartheta$
between the two vectors is conserved in  displacement on a geodesics as shown by  \citep{BouvierM78}, who also  studied 
 geodesics, isometries  and the Killing vectors.

\subsection{Weyl's Integrable Geometry (WIG) and  relation with scalar-tensor theories}  \label{Sweyl}

If a vector is parallelly transported along a closed loop, the total
change of the length of the vector can be expressed as, 
\begin{equation}
\Delta\ell\,=\,\ell\left(\partial_{\nu}\kappa_{\mu}-\partial_{\mu}\kappa_{\nu}\right)\,d\sigma^{\mu\nu}\, ,
\label{boucle}
\end{equation}
where $d\sigma^{\mu\nu}=dx^{\mu}\wedge dx^{\nu}$ is an infinitesimal surface element. 
Weyl identified the tensor $F_{\mu\nu}\,=\,\left(\kappa_{\mu,\nu}-\kappa_{\nu,\mu}\right)$ with the electromagnetic field, 
as its original aim was to also  provide a geometrical interpretation of electromagnetism.
The problem is that this formulation (if the parenthesis does not vanish) implies  non-integrable
lengths so that  the properties of an atom, such as its emission frequencies, would be influenced by its past world line. 
Thus, one could not observe sharp atomic lines  
in the presence of  an electromagnetic field. This was the key point of 
\citet{Einstein18}, who criticized   the use of Weyl's geometry
to describe electromagnetism and gravitation.

In the line of the developments by \citet{Dirac73}, a modified version of Weyl's Geometry  
called Weyl's Integrable Geometry  (WIG) was proposed by \citet{Canuto77}.
In WIG, the above Einstein's remark no longer applies and it may thus  form a consistent framework for
the study of gravitation as emphasized by \citet{Canuto77}, see also \citet{BouvierM78}.
Let us consider a line element
$ds'=g'_{\mu\nu}dx^{\mu}dx^{\nu}$, where the prime symbols apply to  Riemann space, while $ds$ 
 refers to the WIG, which in addition
to the quadratic form  $g_{\mu\nu}$ also has a
scalar gauge field $\lambda(x)$.
In this case,  in Riemann space we have,
\begin{equation}
 \kappa'_{\nu}=0 \, ,
 \label{kz}
 \end{equation}
since  there is no change length in Riemann space.
 According to Eq. (\ref{knu}), we thus  have,
\begin{equation}
\kappa_{\nu}\,=\,-\Phi_{,\nu}\,=\,-\frac{\partial\ln\lambda}{\partial x^{\nu}}\,.
\label{k3}
\end{equation}
This  means that the metrical connexion $\kappa_{\nu}$ in WIG space
is the gradient of a scalar field, 
\begin{equation}
\Phi \, = \,\ln \lambda\, ,
\label{Ph}
\end{equation}
 and  $\kappa_{\nu}dx^{\nu}$ is an exact differential,
\begin{equation} 
\partial_{\nu}\kappa_{\mu}\,=\,\partial_{\mu}\kappa_{\nu} \, .
\label{dx}
\end{equation}
 This  implies that the parallel
displacement of a vector along a closed loop 
does not change its length, see Eq. (\ref{boucle}).
This means that the change of the length  does not
depend on the path followed and thus Einstein's objection does not
hold in  WIG. Nevertheless,  the nice 
 mathematical tools of Weyl's geometry designed for preserving scale covariance (and invariance)
also work in the integrable form of this geometry and this is what we are using in this work. 

At this point, it is appropriate to comment on the relation between the SIV theory and the scalar-tensor theories.
 Alike these theories, but in the context of WIG,  
 SIV enjoys the property of  conformal scale invariance.  In addition, there is also  a coupled scalar field $\Phi$,
 expressed in Eqs. (\ref{k3}) and (\ref{Ph}). However, there is 
a major difference  with  scalar tensor theories. In these,  the scalar field is generally defined in an  independent way, 
it has some specific coupling to standard gravitation, but is not directly determined  by the scale factor. 
  On the contrary, in SIV theory  the scalar field is entierely  defined by the scale factor. Thus,
 there is no degree of freedom to choose the scalar field and the theory is  very constrained.

Now, the question is what is then constraining  the scale factor $\lambda$ in SIV.
 Alike the field equation of GR needs the  specification of the 
metric (Minkowski, Schwarschild, FLWR, etc) characterizing   the physical system under investigation,  SIV needs one more
 specification in the form of a gauging condition, necessary  for fixing the gauge $\lambda$.
\citet{Canuto77} have chosen ``The Large Number Hypothesis''  \citep{Dirac74},
  while  a  different choice is made in Sect. (\ref{fixing}), see also \citet{Maeder17a}.
 
  Many expressions were originally developed 
 by \citet{Weyl23}, \citet{Eddington23}, then extended by \citet{Dirac73} and applied by \citet{Canuto77}
  in  the integrable form of Weyl's Geometry.  On the whole,
WIG  together with  cotensor calculus forms a complete scale covariant framework for gravitation.

\subsection{The geodesic  and the field equations} \label{geodf}

The objective of this work is to show that the Newtonian approximation in SIV  is just leading to the MOND rules  for a valid approximation.
For this purpose we need  the geodesic equation in our  WIG framework.
It has  been  obtained  in different consistent ways. 
 - 1.  It has first been established by \citet{Dirac73} from an action principle.
  - 2. It was obtained  by the covariant generalisation 
  of the usual relation   $u'^{\mu}_{,\nu}\,  u'^{\nu} \, = \,0$ \citep{Canuto77},   
  \begin{equation}
u^{\mu}_{*\nu}\,  u^{\nu} \, = \, 0.
\end{equation}
- 3.  The geodesic in WIG  is  the  curve between two points 
 so that  the change of the length of a vector is minimum:  $\int ^{P_1}_{P_0} \delta(dl)=0$
 \citep{BouvierM78}.
   - 4.  As in GR, the geodesics in WIG   are also  direct consequences 
   of the Equivalence Principle \citep{MBouvier79} in the scale invariant framework,
 \begin{equation}
 \frac{d^2x^{\alpha}}{ds^2}+^*\Gamma^{\alpha}_{\mu  \nu} \frac{dx^{\mu}}{ds} \frac{dx^{\nu}}{ds}+
 \kappa_{\nu}  \frac{dx^{\alpha}}{ds} \frac{dx^{\nu}}{ds}\, = \, 0 \, ,
 \end{equation}
with   $^*\Gamma^{\alpha}_{\mu  \nu}$ given by  Eq.(\ref{eqco}). Thus,  we get
 \begin{equation}
\frac{du^{\alpha}}{ds}+ \Gamma^{\alpha}_{\mu \nu} u^{\mu} u^{\nu} -\kappa_{\mu}u^{\mu} u^{\alpha}+ \kappa^{\alpha} = 0 \, ,
\label{geod}
\end{equation}
\noindent
with the velocity  $u^{\mu} \, = \, dx^{\mu}/ds$.  This expression satisfying the requirement of scale invariance  will be used to derive
 the weak field approximation (Sect. 4.1).  At this stage, the expression of the metrical connexion $\kappa_{\mu}$
is still unknown, since the gauge is not yet fixed. \\

The developments of cotensor calculus of Sect. (\ref{Scotensor}) can be pursued, 
and they are leading to  
a corresponding Riemann--Christoffel tensor  $ ^*R^{\mu}_{\nu \lambda \rho}$ (see Sect. 87  by \citet{Eddington23}; \citet{Dirac73}),
\begin{equation}
^*R^{\mu}_{\nu \lambda \rho}\,  = \, \frac{\partial ^*\Gamma^{\mu}_{\nu \lambda}}{\partial x^{\rho}}+
\frac{\partial ^*\Gamma^{\mu}_{\nu \rho}}{\partial x^{\lambda}}+
 {^*\Gamma^{\eta}_{\nu \lambda}}  {^*\Gamma^{\mu}_{\eta \rho  }}+
{^*\Gamma^{\eta}_{\nu \rho}}{  ^*\Gamma^{\mu}_{\lambda \eta} }   \,.
 \label{Riemcr}
\end{equation}
\noindent
The contracted Riemann--Christoffel tensor  or Ricci tensor appears to be an intensor,  it writes 
in the cotensor form \citep{Eddington23,Dirac73}
\begin{equation}
^*R_{\mu \nu} = R '_{\mu \nu} - \kappa_{\mu, \nu} - \kappa_{ \nu ,\mu}
- g_{\mu \nu}\kappa^{ \alpha}_{,\alpha} -2 \kappa_{\mu} \kappa_{\nu}
+ 2 g_{\mu \nu}\kappa^{ \alpha} \kappa_{ \alpha} \, ,
\label{RC}
\end{equation}
where $R'_{\mu \nu}$ is the usual expression in General Relativity.
The total curvature $R$ in the scale-invariant context is,
\begin{equation}
^*R \, = R^{\alpha}_{\alpha}\,= \,  R' -6 \kappa^{\alpha}_{; \alpha}+6 \kappa^{\alpha} \kappa_{\alpha}, \ 
\label{RRR}
\end{equation}
\noindent
where $R'$ is the total curvature in a standard Riemann geometry. All these expressions,
first and second derivatives, modified Christoffel symbols $^{*}\Gamma_{\mu\nu}^{\alpha}$,
 Riemann--Christoffel tensor $ ^*R^{\mu}_{\nu \lambda \rho}$, 
Ricci tensor and total curvature are scale invariant by construction. The summation and products 
are keeping the cotensoral character and scale covariance. They all contain 
 additional terms depending  on $\kappa_\nu$.  The major difference with these former references rests 
 on the fact that the metrical connexion $\kappa_{\nu}$ is the gradient of the scale factor $\lambda$ 
 by Eq. (\ref{k3}).

 
The scale invariant field equation has also been derived in  different ways \citep{Canuto77}:
 parallelly to GR but with  cotensor expressions taking care of
co-covariant expressions,
 and from an action principle (see below).  In the first method, 
 the first  member $^*R_{\mu  \nu} - \frac{1}{2}  {g_{\mu \nu}} {^*R}$ in the scale invariant context
 is obtained from Eqs. (\ref{RC}) and (\ref{RRR}). Its scale invariant properties are ensured by the cotensoral 
 nature of these expressions.
 
 The second member of the field equation must also be scale invariant, implying that 
 the product of the gravitational constant $G$ and the  energy-momentum tensor $ G   \, T_{\mu \nu}$ 
 should  have the same property.  Unlike   \citet{Dirac74} and   \citet{Canuto77}, $G$ is considered as a constant  
 as in \citet{Maeder17a}.  Thus, one has,
\begin{equation}
 T_{\mu \nu} \,= \,T '_{\mu \nu} \, .
\label{tmn}
\end{equation}
This expression has  implications on the behavior of the relevant densities.
The scale invariance of  tensor $T_{\mu \nu}$  implies,
\begin{equation}
( p+\varrho) u_{\mu} u_{\nu} -g_{\mu \nu } p =
( p'+\varrho') u'_{\mu} u'_{\nu} -g'_{\mu \nu } p' \, .
\end{equation}
There,  velocities $u'^{\mu}$ and $u'_{\mu}$ transform as follows,
\begin{eqnarray}
u'^{\mu}&=&\frac{dx^{\mu}}{ds'}=\lambda^{-1} \frac{dx^{\mu}}{ds}= \lambda^{-1} u^{\mu} \, , \nonumber \\
\mathrm{and} \; \;
u'_{\mu}&=&g'_{\mu \nu} u'^{\nu}=\lambda^2 g_{\mu \nu} \lambda^{-1} u^{\nu} = \lambda \, u_{\mu} \, .
\label{pl1}
\end{eqnarray}
The contravariant and covariant components of a vector have different power, we have seen above that their covariant 
derivatives are different.
Thus, the energy-momentum tensor is scaling like, 
\begin{equation}
( p+\varrho) u_{\mu} u_{\nu} -g_{\mu \nu } p =
( p'+\varrho') \lambda^2 u_{\mu} u_{\nu} - \lambda^2 g_{\mu \nu } p' \, ,
\end{equation}
 implying  the following  behaviour for $p$ and $\varrho$ \citep{Canuto77},
\begin{equation}
p = p' \, \lambda^2 \, \quad \mathrm{and} \quad \varrho = \varrho' \, \lambda^2 \, .
\label{ro2}
\end{equation}
The consistency of the field equation means that pressure and density are not scale invariant, but coscalars of power $\Pi(\rho)=-2$.

The term with the cosmological constant   in Einstein's equation of GR 
is $\Lambda_{\mathrm{E}} \, g'_{\mu \nu}$. According to Eq. (\ref{conformal}), we have
\begin{equation}
\Lambda_{\mathrm{E}} \, g'_{\mu \nu} \, = \, \Lambda_{\mathrm{E}} \lambda^{2}\,g_{\mu\nu}\, \equiv \, \Lambda \, g_{\mu \nu} \, .
\label{Lambda}
\end{equation}
Notation $\Lambda_{E}$
is adopted to avoid any confusion  with  $\Lambda$ in WIG ($\Lambda_{\mathrm{E}}$ 
is not necessarily the value of Einstein static model).
 In the SIV context, $\Lambda= \lambda^2 \Lambda_{\mathrm{E}}$, it is  a coscalar of power $\Pi(\Lambda)=-2$, 
 quite consistently with the previous results for pressure and density. Also, this correspondance will insure the scale iinvariance 
 of the second member of the general field equation.
 Assembling the above expressions developed in the WIG cotensoral   context, we may  express 
the corresponding scale invariant field equation \citep{Canuto77},
\begin{eqnarray}
R'_{\mu \nu} - \frac{1}{2} \ g_{\mu \nu} R'-\kappa_{\mu ,\nu}-\kappa_{ \nu ,\mu}
-2 \kappa_{\mu} \kappa_ {\nu} 
+ 2 g_{\mu \nu} \kappa^{ \alpha}_{,\alpha}
- g_{\mu \nu}\kappa^{ \alpha} \kappa_{ \alpha} = \nonumber  \\
-8 \pi G T_{\mu \nu} - \lambda^2 \Lambda_{\mathrm{E}} \, g_{\mu \nu}, \, 
\label{field}
\end{eqnarray}
\noindent
where $G$ is  a true constant, as seen above.
This is a generalization of Einstein equation in WIG. In addition to the general covariance
 of Einstein equation,  it is  also scale invariant, since  each term  in the equation satisfies this requirement.
The terms with a prime are the same as in GR, the equation
contains additional terms depending on $\kappa_{\nu}$.  This coefficient will be determined by the gauging condition we 
may adopt for the physical system considered, in the same way as the $g_{\mu \nu}$  and their sequence of derivatives (up to the 
Ricci tensor) are determined by the metric,  {\emph{i.e.}} some sort of topography of the space-time system.\\

The above coordinate and  scale invariant field equation (\ref{field}) has also been derived  from an action principle. 
Such a so-called co-covariant equation  has been   developed  by \citet{Dirac73}
for the vacuum  and thus it concerns
    the first member of the general  field equation. As
 this was originally performed  in the  classical Weyl's geometry, the additional field  Dirac introduced in the action
  is independent of the scale factor $\lambda$, in this sense Dirac's results belong to scalar-tensor theories, as he was pointing it
  himself. 

  The study  by \citet{Canuto77} was made  in the line of that by \citet{Dirac73},
  with the difference that it is formally  related  to the WIG framework. The action 
  is a generalization of Einstein action dealing with   
  both coordinate and scale invariant equations,
  
  \begin{equation}
I \, = \, \int\lambda^{2}\left(*R\right)\sqrt{g} \,d^{4}x \, .
\label{act1}
\end{equation}

\noindent 
  The scale factor $\lambda$ is a coscalar of power $\Pi(\lambda)=-1$,
while $\left(*R\right)$ (so noted as a coscalar) has a power $\Pi(*R)=-2$,
 since it is a contraction of the intensor $R_{\mu\nu}$. We also have
 $\Pi(g_{\mu\nu})=2$, and thus  $g=\det(g_{\mu\nu})$
is  a co-scalar of power $\Pi(g)=8$. 
The multiplication by $\lambda^2$ garantees the invariant property. Terms being 
 functions of $\lambda$ and of  its cotensor derivatives  are added to the above equation, no  other new  field is introduced.
 The action principle writes,
 
\begin{equation}
\delta I = \delta \int \left( -\lambda^{2} \, *R+ c_1 \lambda^{*\mu} \lambda_{* \mu}+ c_2  \lambda^4 \right)\sqrt{g} \,d^{4}x=0 \, ,
\label{act2}
\end{equation}

\noindent
where $c_1$ and $c_2$  are constants.
The quartic term then is related to the cosmological constant.
A matter Lagrangian $\mathcal{L}$ may be included, its relation with the energy momentum tensor is,
\begin{equation}
G \, T^{\mu \nu} \, = \, \lambda^{-2} \, \frac{2}{\sqrt{g}} \frac{\delta  \sqrt{g}\; \mathcal{L}}{\delta g_{\mu \nu}} \,,
\label{act3}
\end{equation}
with the term $\lambda^{-2}$ for power  consistency, while
 the Lagrangian density $\mathcal{L}$ must be an inscalar expression.
 As a result, the development of the above expressions  confirms the above scale invariant field equation (\ref{field})
  and,  consistently enough,  produces no new field equations.

 \subsection{Fixing the gauge}   \label{fixing}
 
Since scale covariance is considered in addition to the coordinate covariance of GR, an 
  additional condition   is  necessary to specify the gauge $\lambda$. 
  \citet{Dirac73} and  \citet{Canuto77}  had chosen the  so-called {\it{``Large
Number Hypothesis``}}, see also \citet{Dirac74}. 
The author's choice is to adopt   the following statement \citep{Maeder17a}:
  {\emph{The macroscopic empty space is scale invariant, homogeneous and isotropic}}.
 This   assumption  is  consistent with the scale invariance of GR in empty space and with Maxwell's
 equations in absence of charges and currents. 
  Moreover, the  equation of state of the vacuum $p_{\mathrm{vac}}= -\varrho_{\mathrm{vac}}  c^2$  is precisely 
the relationship permitting  the vacuum density   to remain constant for an adiabatic
expansion or contraction  \citep{Carroll92}.

Under the above key hypothesis, 
one is left from the general  field Eq.(\ref{field}) with the following condition for the empty spacetime, 
\begin{equation}
\kappa_{\mu;\nu}+\kappa_{\nu;\mu}+2\kappa_{\mu}\kappa_{\nu}
-2g_{\mu\nu}\kappa_{;\alpha}^{\alpha}+g_{\mu\nu}\kappa^{\alpha}\kappa_{\alpha}
=\lambda^{2}\Lambda_{\mathrm{E}}\,g_{\mu\nu}.\label{fcourt}
\end{equation}
\noindent
The geometrical terms $R'_{\mu\nu}$ and $R'$ of the field equation have disappeared, 
since the de Sitter metric for an empty space endowed with a cosmological constant 
is conformal to the Minkowski metric where $R'_{\mu\nu}=0$ and $R'=0$. 
The conformal relation becomes an identity,   if $3 \lambda^{-2}/(\Lambda_{\mathrm{E}} t^2)=1$ \citep{Maeder17a}, a condition 
which is  noticeably fully consistent with the solution of (\ref{fcourt}), 
as shown below.

We assume that the macroscopic empty space characterized by the above equation is 
homogeneous and isoptropic. This is consistent with the hypothesis that the scale factor $\lambda$  is a function of time only.
Thus, only the zero component of $\kappa_\mu$ is non-vanishing and the coefficient of metrical connection becomes,
\begin{eqnarray}
\kappa_{\mu ;\nu} = \kappa_{0 ;0} = \partial_0 \kappa_0 = \frac{d \kappa_0}{dt} \equiv
{\dot{\kappa}_0} \, =-\frac{\dot{\lambda}}{\lambda} \,,
\label{k0}
\end{eqnarray}
\noindent
The 0 and the 1, 2, 3 components of what remains from Eq. (\ref{fcourt}) become respectively:
\begin{equation}
3 \kappa^2_0 \, = \,\lambda^2 \, \Lambda_{\mathrm{E}} \, , \quad \mathrm{and}   \; \; 2 \dot{\kappa}_0 - \kappa_0^2 = -\lambda^2 \Lambda_{\mathrm{E}} \, .
\label{k1}
\end{equation}
\noindent
and we get the two most important equations,
\begin{eqnarray}
\  3 \, \frac{ \dot{\lambda}^2}{\lambda^2} \, =\, \lambda^2 \,\Lambda_{\mathrm{E}}  \,  
\quad \mathrm{and} \quad  \,  2\frac{\ddot{\lambda}}{\lambda} - \frac{ \dot{\lambda}^2}{\lambda^2} \, =
\, \lambda^2 \,\Lambda_{\mathrm{E}}\,,
\label{diff1}
\end{eqnarray}
\noindent
or some combinations of them. These two expressions have important consequences:

 - In GR, $\Lambda_{\mathrm{E}}$ and the properties of the empty space are considered not to depend on the matter content 
 of the Universe. The same applies to the above two equations and to  the scale factor $\lambda$.
 This  is also consistent with the fact that matter density does not appear in these two equations.

- These differential equations establish a relation of the cosmological constant, or the energy density of the vacuum,
 with the scale factor $\lambda$ and its variations.

 - From the relation $\Lambda = 8 \pi G \varrho_{\mathrm{vac}}$ 
 and the first of the Eqs. (\ref{diff1}), the energy 
density of the vacuum  can  be expressed in term of a scalar field $\psi$,
 \begin{equation}
\varrho \, = \, \frac{1}{2}\, C \,  \dot{\psi}^2  \, \quad \mathrm{with}   
\; \,\dot{\psi}  \,= \,  \kappa_0 = \,- \frac{\dot{\lambda}}{\lambda} \, .
\label{ro2}
\end{equation}
with  constant $C= 3/(4\pi G)$. The field $\psi$ obeys a modified Klein-Gordon equation \citep{MaedGueor21a}.
 
 - For a  solution of Eqs. (\ref{diff1})  of the form $\lambda= a (t-b)^n + d$, 
 we get $d=0$,  $n=-1$ with $a= \sqrt{\frac{3}{\Lambda_{\mathrm{E}}}}$. 
 There is no condition on $b$ from Eqs.(\ref{diff1}), any value would fit. 
 However, the solutions of the cosmological equations  may put some conditions on the origin of time, 
 depending on model  parameters, see Sect. 3.
 Thus, with this remark the general solution is, 
\begin{equation}
\lambda(t) =\sqrt{\frac{3}{\Lambda_{\mathrm{E}}}}  \frac{1}{ct} \, .
\label{lambda}
\end{equation}
 Thus, if we adopt the scale factor $\lambda_0=1$ at the present time $t_0=1$,  we just have 
$\lambda(t)= (t_0/t)$ in a system of units where $\sqrt{\frac{3}{c^2 \,\Lambda_{\mathrm{E}}}}=1$.

 
\section{Cosmological solutions and their implications}   \label{equ}

The application of the  FLWR  metric to Eq.(\ref{field})  leads to 
 cosmological  equations  \citep{Canuto77},
\begin{eqnarray}
\frac{8 \, \pi G \varrho }{3} = \frac{k}{a^2}+
\frac{\dot{a}^2}{a^2}+ 2 \, \frac{\dot{\lambda} \, \dot{a}}{\lambda \, a}+
\frac{\dot{\lambda}^2}{\lambda^2} - \frac {\Lambda_{\mathrm{E}} \lambda^2}{3} \, ,  \label{E1p}  \\
-8 \, \pi G p = \frac{k}{a^2}+ 2 \frac{\ddot{a}}{a} + 2 \frac{\ddot{\lambda}}{\lambda}+
\frac{\dot{a}}{a}^2+ 4 \frac{\dot{a} \, \dot{\lambda}}
{a \, \lambda}-\frac{\dot{\lambda^2}}{\lambda^2} -\Lambda_{\mathrm{E}} \,  \lambda^2  \, ,\label{E2p}\\
\mathrm{with} \; \;\varrho \, a^{3(1+c^2_s)} \lambda^{1+3c^2_s} = \mathrm{const.} \label{conserv}
\end{eqnarray}
\noindent
The last expression is the  conservation law,  with a sound velocity $c^2_s =0$
for a dust model and $c^2_s=1/3$ for the radiative era.  If  $\lambda$ is a constant, the derivatives of $\lambda$
  vanish  and one is brought back to the equations of GR.
   Solutions of these equations have been searched by \citet{Canuto77} for two variant cases of the Large Number
  Hypothesis.  The major point is that the above equations are leading to  solutions characterized by  expansion 
    factors $ a(t) \, \sim  \, t$ at large cosmological times, a situation no longer supported nowadays.

Remarkably, 
the gauging condition, which implies  Eqs. (\ref{diff1}), 
leads  to major simplifications of Eqs. (\ref{E1p}) and  (\ref{E2p}), which become \citep{Maeder17a},
\begin{eqnarray}
\frac{8 \, \pi G \varrho }{3} = \frac{k}{a^2}+\frac{\dot{a}^2}{a^2}+ 2 \,\frac{\dot{a} \dot{\lambda}}{a \lambda} \, ,
\label{E1} \\
-8 \, \pi G p  = \frac{k}{a^2}+ 2 \frac{\ddot{a}}{a}+\frac{\dot{a^2}}{a^2}
+ 4 \frac{\dot{a} \dot{\lambda}}{a \lambda}  \, .
\label{E2}
\end{eqnarray}
\noindent
For a constant $\lambda$, Friedmann's equations are recovered.
 A  third equation may be derived from the above two,  
 \noindent
\begin{equation}
- \frac{4\pi G}{3} \left(3p+\varrho \right) = \frac{\ddot{a}}{a} + \frac{\dot{a} \dot{\lambda}}{a \lambda} \, .
\label{E3n}
\end{equation}
\noindent
Since $\dot{\lambda}/{ \lambda}$ is negative,  the extra term represents  an additional acceleration {\emph{ in the direction of  
motion}}. This  effect of the scale invariance is fundamentally different from that of the cosmological constant.
Now, for an expanding Universe, this extra force produces an accelerated expansion, without requiring 
 dark energy particles.
For  a contraction, 
 the additional  term   favours collapse, as shown
  in the study of the growth of density fluctuations \citep{MaedGueor19}, where  
 an early formation of galaxies  is resulting without the need of dark matter.\\
 
 \begin{figure}
\centering
\includegraphics[angle=0.0,height=5.7cm,width=8.2cm]{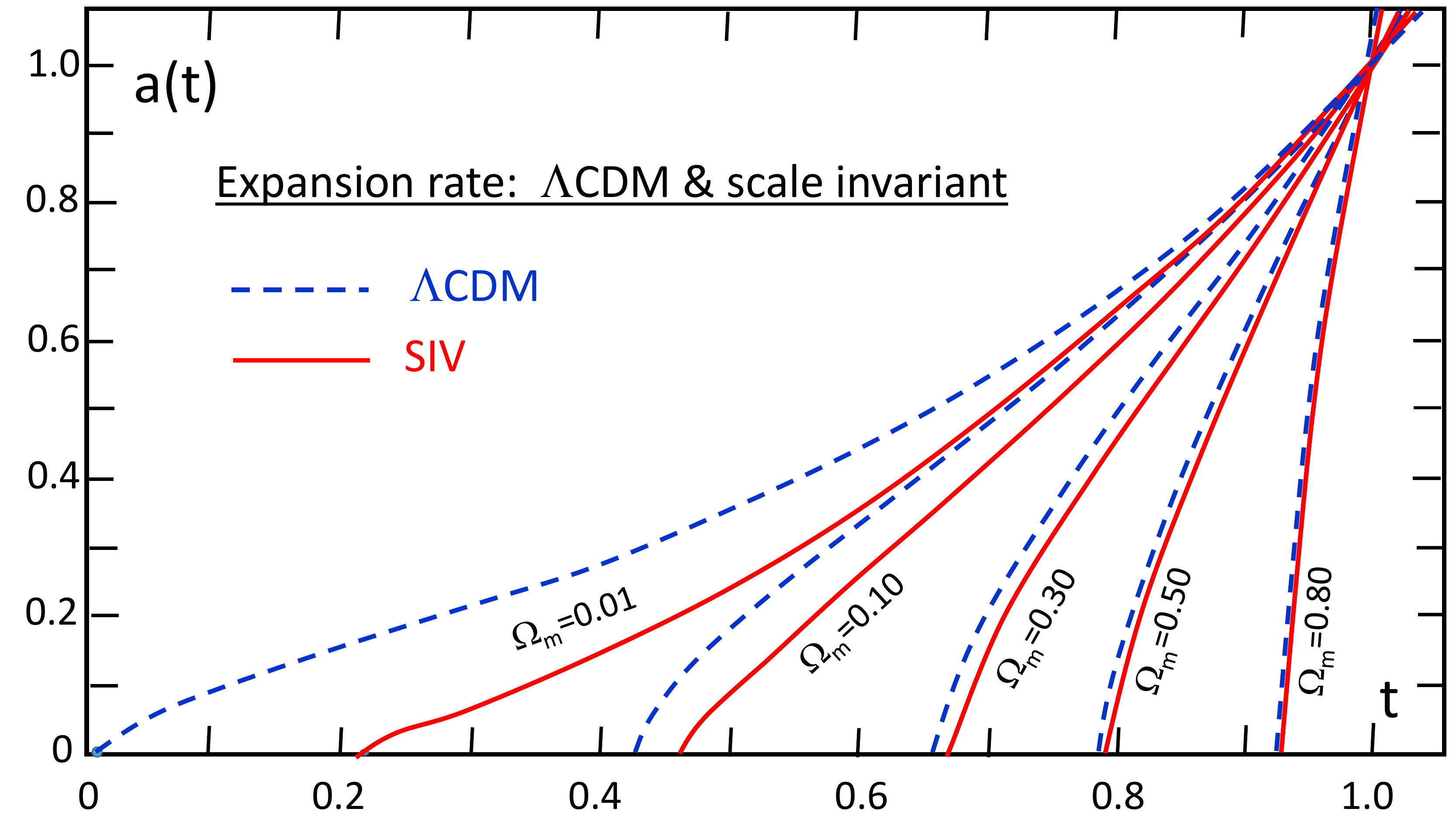}
\includegraphics[angle=0.0,height=6.2cm,width=8.5cm]{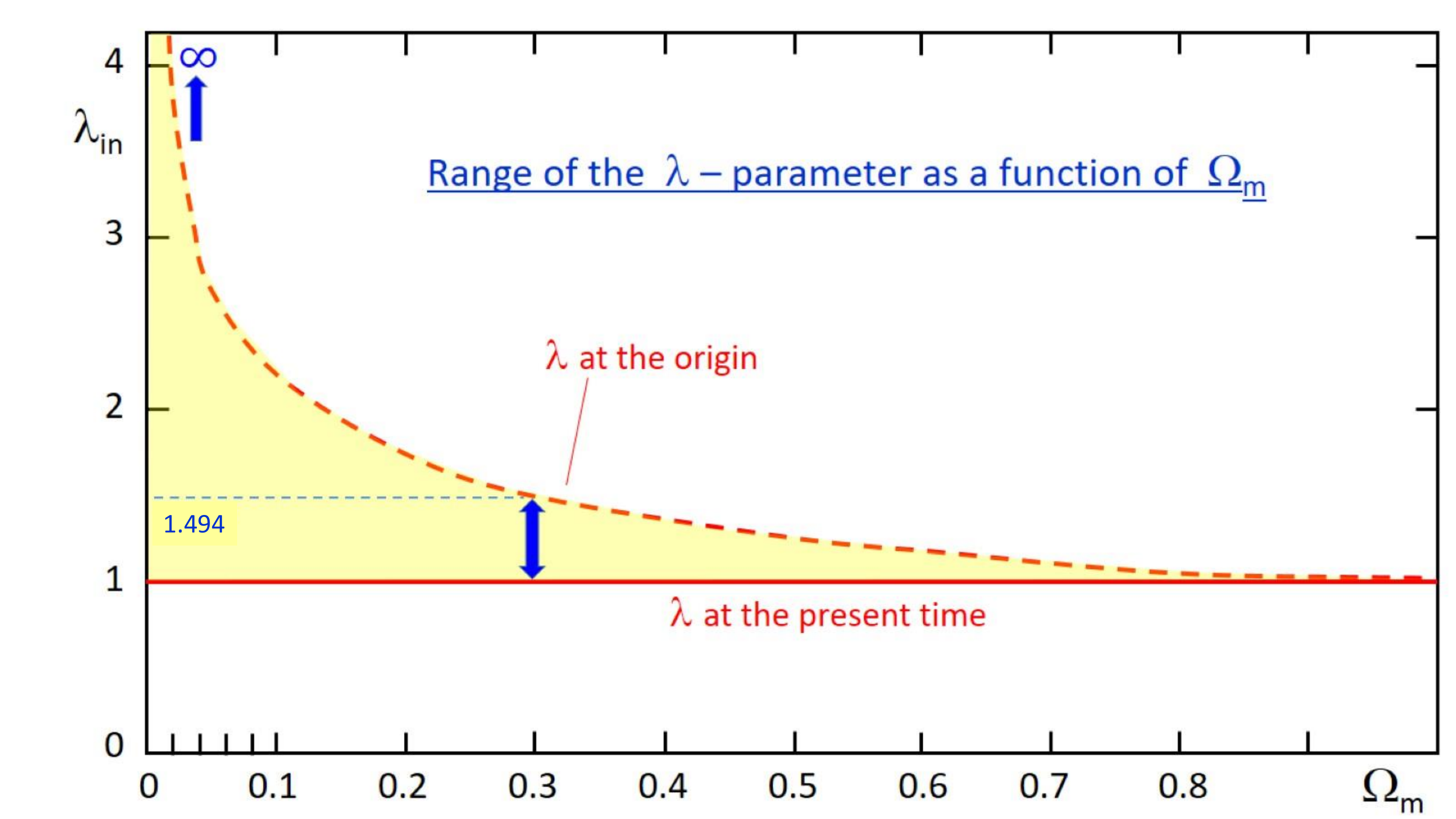}
\caption{Top: The expansion factor $a(t)$  for the $\Lambda$CDM  and
SIV models as a function of  $\Omega_{\mathrm{m}}$. 
Bottom: The  scale factor  $\lambda_{\mathrm{in}}= 1/t_{\mathrm{in}}$ at the initial time
 $t_{\mathrm{in}}=\Omega^{1/3}_{\mathrm{in}}$   
as a function of $\Omega_{\mathrm{m}}$. 
The yellow  zone shows, vs. $\Omega_{\mathrm{m}}$,  the range of  $\lambda(t)$  
  from the Big-Bang (broken red line) to the present time (continuous red line).
For $\Omega_{\mathrm{m}}=0.3$,  $\lambda(t)$  only varies  between 1.494 at the origin 
and 1.0 at present.}
\label{courbes}
\end{figure}
 
 The solutions of these equations have been discussed in details in \citet{Maeder17a}, together 
 with various cosmological properties concerning the Hubble-Lema\^{i}tre and deceleration  parameters, the 
 cosmological distances and  different
 cosmological tests. The redshift drifts appear as one of the most promising cosmological tests  \citep{MaedGueor20a}.
 Here, we limit the discussion to a few points pertinent to the subject of the paper.
Analytical solutions  for the flat SIV models with $k=0$,  considered here,  have been found 
for the  matter  \citep{Jesus18} and radiation  \citep{Maeder19} dominated  models. In the former case, we have
\begin{equation}
a(t) \, = \, \left[\frac{t^3 -\Omega_{\mathrm{m}}}{1 - \Omega_{\mathrm{m}}} \right]^{2/3}\, .
\label{Jesus}
\end{equation}
\noindent
It is expressed in the timescale $t$ where at present  $t_0=1$ and  $a(t_0)=1$. Such solutions are illustrated in Fig. \ref{courbes}, top.
They are lying relatively close to the $\Lambda$CDM ones, the differences being larger for lower $\Omega_{\mathrm{m}}$.
 This is a general property: {\it{the effects of scale invariance are always larger for the lower matter densities, being the
largest ones for the empty space. }}
There $\Omega_{\mathrm{m}}= \varrho/\varrho_{\mathrm{c}}$  with $\varrho_{\mathrm{c}}=3H^2_0/(8\pi G)$.
 Remarkably, Eqs. (\ref{E1}) and (\ref{E2}) allow flatness for different values of  $\Omega_{\mathrm{m}}$, unlike the classical
 Friedmann models.
 
The initial time when $a( t_{\mathrm{in}})=0$ is,

\begin{equation}
 t_{\mathrm{in}} \, = \, \Omega^{1/3}_{\mathrm{m}}\,.
 \label{tin}
 \end{equation}
 This dependence in 1/3 produces a  rapid increase of $ t_{\mathrm{in}}$   for increasing  $\Omega_{\mathrm{m}}$.
 For $\Omega_{\mathrm{m}}=0, 0.01, 0.1, 0.3, 0.5$, the values of  $ t_{\mathrm{in}}$ are 0, 0.215, 0.464, 0.669, 0.794 respectively.
 The key point is that this leads to a strong reduction of the range of $\lambda(t)$ for increasing $\Omega_{\mathrm{m}}$
 (Fig. 1, bottom):  while the range of $\lambda$ is infinite  for an empty model,
 it is very limited for significant $\Omega_{\mathrm{m}}$-values. 
  Thus,  the presence of matter  through $\Omega_{\mathrm{m}}$ drastically  reduces  
  the range of variation of  the  universal scale factor $\lambda$.  For $\Omega_{\mathrm{m}} > 1$
  scale invariance is killed, which   makes sense in view of the remarks by \citet{Feynman63}. 
    This is a global effect  associated to the range of $\lambda$ 
  in  Universe  models.
  This does not prevent other effects due to local variations of density to also  intervene, as shown in Sect. (\ref{2ap}), see Eq. (\ref{x2}). 
 

 The Hubble parameter is ,  in the  timescale $t$ (which goes from $t_{\mathrm{in}}$ at 
  Big-Bang to $t_0=1$ at present),
 \begin{equation}
 H(t) = \frac{2 \, t^2}{t^3- \Omega_{\mathrm{m}}}\,.
 \label{HM}
 \end{equation}
From Eqs. (\ref{Jesus}) and (\ref{HM}), we  see that there is no meaningful scale invariant solution
 for an expanding Universe model with
$\Omega_{\mathrm{m}}$ equal or larger than 1.
Thus, the model solutions   are  quite  consistent with  the  causality relations  discussed by \citet{MaedGueor21a}.

One can also define  
\begin{equation}
\Omega_{\mathrm{k}} = -\frac{k}{a^2  H_0^2} \, \quad \mathrm{and} 
\quad \Omega_{\lambda} \, =\,- \frac{2}{H_0} \left(\frac{\dot{\lambda}}{\lambda}\right)_0 \,= \,\frac{2}{ H_0 \, t_0} \,.
\label{hl}
\end{equation}
\noindent
These are  the normalized  contributions  vs. $\varrho_{\mathrm{c}}$ respectively
of the matter, space curvature, and scale factor $\lambda$.
With these definitions,  the first cosmological equation  (\ref{E1}) leads to, 
\begin {equation}
\Omega_{\mathrm{m}} \, + \, \Omega_{\mathrm{k}} \, +  \Omega_{\lambda} = \, 1  \, .
\label{Omegapr}
\end{equation}
\noindent
These quantities are usually considered
at the present time. 

 \section{The Newtonian and MOND approximations}   \label{Newton}
 
 \subsection{The weak field approximation}
 The scale invariant expression of the geodesic equation was derived by \citet{Dirac73} and also from an action principle
 by \citet{BouvierM78}, see Sect. \ref{geodf}.
The weak field low velocity approximation has been obtained
by \citet{MBouvier79}, see also \citet{Maeder17c},
 \begin{equation}
 \frac {d^2 \bf{r}}{dt^2} \, = \, - \frac{ G_t \, M(t)}{r^2} \, \frac{\bf{r}}{r}   +  \kappa(t) \,\frac{d\bf{r}}{dt} \, ,
\label{Nvec}
\end{equation}
 in spherical coordinates. It contains an additional acceleration term in the direction of motion,  {\it{the dynamical gravity}}. 
This term, proportional to the velocity,  favours collapse during a contraction,   and outwards motion if expansion.

The conservation law  (Eq. \ref{conserv}) imposes  for a dust Universe a relation  $\varrho a^3 \lambda=const.$,  %
meaning that the inertial mass of a particle is not a constant and that it depends on the scale factor $\lambda$. 
  We note
 that the non-constancy of mass is also a common situation in Special Relativity. 
Here,  masses  vary  like $M(t)  =   M(t_0)  (t/t_0)$. Interestingly enough, 
rather than the  inertial and gravitational mass, the gravitational potential  
 $\Phi =G\, M/r$ of an object, thus the field, appears as a more fundamental quantity, 
 being    scale-invariant through the evolution of Universe. As an example, 
for $\Omega_{\mathrm{m}} =0.3$, the mass  at the Big-Bang was,
  $M(t_{\mathrm{in}}) =  \Omega_{\mathrm{m}}^{1/3} \, M(t_0)= 0.6694 \; M(t_0)$, 
the variations are  smaller than 1\% over the last 400 million years (Sect.\ref{M1}).

 
 In the above cosmological models, the age $t$  is $t_0=1$ at  present 
 and  $t_{\mathrm{in}} = \Omega^{1/3}_{\mathrm{m}}$ at the origin. 
 The usual timescale $\tau$  in years or seconds  is $\tau_0= 13.8$ Gyr at  present  \citep{Frie08} 
 and  $\tau_{\mathrm{in }}=0$ at the Big-Bang.
 Thus, the  relation between these  ages   is,
\begin{equation}
\frac{\tau - \tau_{\mathrm{in}}}{\tau_0 - \tau_{\mathrm{in}}} = \frac{t - t_{\mathrm{in}}}{t_0 - t_{\mathrm{in}}}\, ,
\end{equation}
expressing that the age fraction with respect to the present age is the same in both timescales.
This gives 
\begin{equation}
\tau \,= \, \tau_0 \, \frac{t- \Omega^{1/3}_{\mathrm{m}}}{1- \Omega^{1/3}_{\mathrm{m}}} \,  \quad \mathrm{and} \; \;
  t \,= \, \Omega^{1/3}_{\mathrm{m}} + \frac{\tau}{\tau_0} (1- \Omega^{1/3}_{\mathrm{m}}) \,,
\label{T2}
\end{equation}
and for the derivatives,  
\begin{equation}
\frac{d\tau}{dt} \, = \, \frac{\tau_0}{1-\Omega^{1/3}_{\mathrm{m}}}\,, \quad \mathrm{and}\; \;
\frac{dt}{d\tau} \, = \, \frac{1-\Omega^{1/3}_{\mathrm{m}}}{\tau_0}\,.
\label{dT1}
\end{equation}
For larger  $\Omega_{\mathrm{m}}$,   timescale $t$ is squeezed over a smaller fraction of the interval
0 to 1.0, (which reduces the range of $\lambda$ over the ages). \\

 We need to convert the  equation of motion (\ref{Nvec}) expressed with variable $t$ into the usual time $\tau$. 
Equation (\ref{Nvec})  becomes,
 \begin{equation}
 \frac {d^2 \bf{r}}{d \tau^2} \left(\frac{d \tau}{dt}\right)^2 \, = \, - \frac{G_t \, M(t)}{r^2} \, \frac{\bf{r}}{r}   +
  \, \frac{1}{ t_\mathrm{in} +   \frac{\tau}{\tau_0} (t_0-t_\mathrm{in}) } \, \frac{d \tau}{dt}  \frac{d\bf{r}}{d\tau} \, \, .
\label{Nvec2}
\end{equation}
Here $G_t$ is used to specify that  the  gravitational constant  is expressed with time units $t$.
  In the  $\tau$-scale, the  units of $G$ are $ [cm^3\cdot g^{-1} s^{-2}]$,  thus,  the correspondence is
$G_t \,  \left(\frac{dt}{d\tau}\right)^2 = G$. At  present,  
the masses $M(t_0)$ and  $M(\tau_0)$ are evidently equal. At other epochs, the relation is,
\begin{equation}
M(t)= \frac{t}{t_0}  M(t_0),  \;   \mathrm{thus} \;
M(\tau)=\left[ \Omega^{1/3}_{\mathrm{m}}+\frac{\tau}{\tau_0} (1- \Omega^{1/3}_{\mathrm{m}}) \right]  M(\tau_0).
\label{mass}
\end{equation}
\noindent
Now, multiplying both members of Eq.(\ref{Nvec2}) by $\left(\frac{dt}{d \tau}\right)^2$, we get at time  $\tau/\tau_0$,
 \begin{equation}
 \frac {d^2 \bf{r}}{d \tau^2}  \, = \, - \frac{G \, M(\tau)}{r^2} \, \frac{\bf{r}}{r}   
+ \, \frac{1}{ t_\mathrm{in} +   \frac{\tau}{\tau_0} (t_0-t_\mathrm{in}) } \, \frac{t_0-t_\mathrm{in}}{\tau_0}\,\frac{d\bf{r}}{d\tau} \, \, .
\label{Nvec3}
\end{equation}
We define the numerical factor $\psi$,
\begin{equation}
\psi(\tau) \, = \, 
\frac{t_0-t_\mathrm{in}}{ t_\mathrm{in} +   \frac{\tau}{\tau_0} (t_0-t_\mathrm{in}) }\, ; \; \mathrm{thus} \;
\psi_0=\psi(\tau_0) \, =1-\Omega^{1/3}_{\mathrm{m}} \,.
\label{phi}
\end{equation} 
The modified Newton's equation at present time $\tau_0$ is then,
\begin{equation}
 \frac {d^2 \bf{r}}{d \tau^2}  \, = \, - \frac{G \, M(\tau_0)}{r^2} \, \frac{\bf{r}}{r}   + \frac{\psi_0}{\tau_0}   \frac{d\bf{r}}{d\tau} \, \,.  
\label{Nvec4}
\end{equation}
\noindent
The  additional  term, {\it{the dynamical gravity}}, which is  generally extremely small (cf. Eq. \ref{x2}),
 also depends  on  $\Omega_{\mathrm{m}}$:  
 in an  empty Universe, $\psi_0= 1$, the effect being maximum,  while  for  $\Omega_{\mathrm{m}} =1$, one consistently has $\psi_0 =0$,
scale invariance has no effect.
For $\Omega_{\mathrm{m}} = 0.30$, 0.20, 0.10 and 0.05  one has $\psi_0=0.331, 0.415, 0.536$  and 0.632,
 which reduces  the dynamical gravity.

  \subsection{The MOND approximation: first approach} \label{M1}
  
  Let us first examine how the scale factor $\lambda(\tau)$ and consequently the  masses $ M(\tau)$ are varying over  the past. 
  For the case $\Omega_{\mathrm{m}}=0.30$
  as an example, over the last 100 Myr, 200 Myr and  0.5 Gyr the   mass increase  predicted by  Eq. (\ref{mass}) amounts to a factor 1.0024,
  1.0048 and  1.012 respectively. 
  For $\Omega_{\mathrm{m}}=0.10$,  these values would be 1.0039, 1.0078 and 1.020. 
  
  Thus, for galaxies where the 
  rotation periods are  a few hundered millions years, we can consider 
  that both $\lambda$ and   masses are constant with an accuracy equal or better than 1\%.  
  Indeed, this is quite consistent with MOND, which is also known to be scale invariant  with a constant scale factor $\lambda$
  \citep{Milgrom15}.
  Such an approximation is much less satisfactory for clusters of galaxies where the time scales, {\it{e.g.}} the crossing times,
  are of the order of a few Gyr.  
  
  With a constant $\lambda$, the coefficient  $\kappa= -\frac{\dot{\lambda}}{\lambda}$ 
  is equal to zero  and  the   dynamical gravity  in Eq. (\ref{Nvec4})
  disappears. We are left with transformations $r = \lambda \; r' $  and  $t = \lambda \; t'$ with a constant $\lambda$ (and thus $M$)
    applied to the Newton equation expressed  in the prime coordinates,
   \begin{equation}
     \frac{d^2r'}{dt'^2} \, = \, - \frac{GM}{r'{^2}} \,  \equiv  \, g'_N.
     \end{equation}
 The total acceleration $g$ in  system  $(r, t)$ becomes,
 \begin{equation}
 g\, = \, \frac{d^2r}{dt^2} \,=\,\frac{1}{\lambda} \, \frac{d^2 r'}{dt'^{2}}\,. 
 \label{e1}
 \end{equation}
 \noindent
 The  Newtonian gravitational accelerations $g_N$ and $g'_N$ are related  by,
 \begin{equation}
g_N \, \equiv \, - \frac{G \, M}{r^2} \, = \,- \frac{1}{\lambda^2}  \frac{G \, M}{r'^2}\, \equiv   \frac{1}{\lambda^2} \,g'_N\, .
 \label{e2}
 \end{equation}
 \noindent
 Eq. (\ref{e1})  can thus be developed as follows,
 \begin{equation}
 g \, = \,\frac{d^2r}{dt^2} \,=\,\frac{1}{\lambda} \, \frac{d^2r'}{dt'^2}\, = \, \frac{1}{\lambda} \, g'_N \, = \, {\lambda} \, g_N \, ,
 \label{e3}
 \end{equation}
 according  to (\ref{e2}). This last relation also implies,
 \begin{equation}
g \, = \,  \frac{d^2r}{dt^2} \,= \, {\lambda} \, g_N \,  = \, \left(\frac{ g'_N}{g_N}\right)^{1/2}g_N \, = \left(g'_N \, g_N \right) ^{1/2} \, .
 \label{e4}
 \end{equation}
 This is to be compared to  the deep-MOND limit given by Eq. (\ref{deep}). 
 We note a correspondence between  constant $a_0$ and  $g'_N$.
  At this stage,   we have no information on what kind of value should  be used for $g'_N$, and for what range of gravities 
  it may apply.  In a second approach
 below, we will get more information on these points. For now, we note that the approximation of constant $\lambda$ 
 and masses over 
 a few hundred millions years in the scale invariant theory just leads to a form analog to the deep-MOND limit.
 
 The  constant   $a_0$ is related  to  $c \, H_0$  \citep{Milgrom83,Milgrom15}.  We may  wonder why
 a constant, considered as a universal constant,  thus being the same at any time, 
 should just be related to the present value $H_0$, see also \citep{Milgrom20}. 
 This is highly suggestive   that $a_0$, alike $\lambda$, is in fact  time dependent, contrarily to MOND assumption,
 but in agreement with the scale invariant theory.

 \subsection{The MOND approximation: second approach}  \label{2ap}
 
 We may also derive the  MOND behaviour at very low gravities  from   the equation of motion 
 Eq. (\ref{Nvec4}).
 The ratio  $x$ of the radial components of the  dynamical gravity to the Newtonian one   is given by,
 \begin{equation}
 x\, = \, \frac{\psi_0 \, \upsilon r^2} {\tau_0 \, GM}\,.
 \label{xx}
 \end{equation}
 \noindent
 where $\upsilon$ is the radial component of the velocity. 
 The  ratio $x$ may become larger than 1 in two  particular cases: 
 
 - 1. In very early stages of the
 Universe,  $\tau_0$ is very small and favours  a large dynamical acceleration $(\psi \, \upsilon)/\tau_0$ 
 in the sense of motion. This effect is likely to have favoured the early galaxy formation without the need of
 dark matter \citep{MaedGueor19}. 
 
 - 2. At  large distances from a central
 body,  the Newtonian gravity $g_N$ may become smaller than  $(\psi \, \upsilon)/\tau_0$. This may typically occur 
 in very wide binaries and in the  outer layers of galaxies and clusters.
 This situation is  favoured by the fact that  both the  deep-MOND Limit and   SIV theory predict that
 the  orbital velocities in  two-body systems  are independant of the orbital radius \citep{Milgrom14b,MBouvier79}.
   Thus, in both cases  larger orbital velocities 
 may favour the dynamical acceleration  $(\psi \, \upsilon)/\tau_0$  in  the outer layers of gravitational systems.\\

 We use the fact that $H_0= \frac{\xi}{\tau_0}$  and  $\varrho_{\mathrm{c}}= \frac{3 \, H_0^2}{8 \, \pi \,G}$ to express 
 $\tau_0$ in term of $\varrho_{\mathrm{c}}$ in Eq.(\ref{xx}). 
 To obtain $\xi$, we use the following expression of the Hubble-Lema\^{i}tre parameter (see Appendix),
\begin{eqnarray}
 H(\tau_0) \,= \,\frac{ 2}{1- \Omega_{\mathrm{m}}} \, \frac{(1-\Omega^{1/3}_{\mathrm{m}})}{\tau_0} \, ,      \\
   \quad  \mathrm{thus} \; \; \xi\,=\, \frac{ 2 \, (1-\Omega^{1/3}_{\mathrm{m}}) }{(1- \Omega_{\mathrm{m}})}.  
 \label{hx}
   \end{eqnarray}
 The ratio  $\xi$ is about unity in the SIV theory
 for $\Omega_{\mathrm{m}}=0.10, 0.20$ or 0.30,  there one has $\xi =1.191, 1.038$ or 0.945 respectively. 
 We  consider a mass $M$  spherically distributed in a radius $r$ with a mean density $\varrho$ and get, 
 \begin{equation}
 x \, = \, \frac{\sqrt{2} \, \psi_0}{\xi}  \left(\frac{\upsilon^2}{(GM/r)}\frac{\varrho_{\mathrm{c}}}{\varrho} \right)^{1/2}  \, .
 \end{equation}
 \noindent
 Let us consider a two-body system formed by a massive central mass $M$ and a test object of negligible mass, 
 there  the instantaneous equilibrium of forces determined by Eq. (\ref{Nvec4}) along the radial direction
 is just  $\frac{\upsilon^2}{r} = \frac{GM}{r^2}$, since the additional dynamical acceleration is in the direction of the
 motion\footnote{When one considers the average velocity dispersion in a long-time evolution, this type of 
 relation does not necessarily   hold any more \citep{Maeder17c})}. Thus, in external regions of galaxies where normally
 the motion should be mainly determined by the central mass concentration, the ratio $x$ becomes,
 \begin{equation}
  x \, = \, \frac{\sqrt{2} \, \psi_0}{\xi} \left( \frac{\varrho_{\mathrm{c}}}{\varrho} \right)^{1/2}  \, . 
  \label{x2}
 \end{equation}
 For high values of the density $\varrho$ with  respect  to the critical density,
  the $x$-parameter  becomes negligible and thus 
 the gravity is just determined by the usual Newton Law.
 \noindent
 At the edge of the sphere of radius $r$ and mean density $\varrho$, the Newtonian gravity  $g_N= (4/3) \pi G \, \varrho \, r$, so that
  we also  have,
 \begin{equation}
  x \, = \, \frac{\sqrt{2}\,  \psi_0}{\xi} \left( \frac{g_{\mathrm{c}}}{g_{\mathrm{N}}} \right)^{1/2}  \, ,
 \end{equation}
 \noindent
 where $g_{\mathrm{c}}$ is the mean gravity at the edge of a similar sphere having the critical density.
 Thus, the total gravity $g$ given by the the first member of Eq.(\ref{Nvec4}) is
 \begin{equation}
 g \, =  \, g_{\mathrm{N}} + x \, g_{\mathrm{N}} \,= g_{\mathrm{N}}\left[1+  \frac{\sqrt{2}\,  \psi_0}{\xi} \,
 \left(\frac{g_{\mathrm{c}}}{g_{\mathrm{N}}}\right)^{1/2} \right]\,.
 \end{equation}
 \noindent
 Let us consider regions at large  distances from the galactic center, or in a binary system at large distances from the other body.
 There, the   Newtonian gravity   $g_{\mathrm{N}}$, resulting from the attraction of the  central or other body, 
 can be counterbalanced by  the gravity  of  external galaxies or other star systems so that the resulting  $g_{\mathrm{N}}$
 essentially vanishes and  $x$ becomes bigger  than 1 in large regions. 
 When this  happens, we get  a total resulting gravity $g$ behaving like,
  \begin{equation}
 g \, \rightarrow  \,  \frac{\sqrt{2}\,  \psi_0}{\xi} \, \left( g_{\mathrm{c}} \, g_{\mathrm{N}} \right)^{1/2}\,.
 \label{g1}
 \end{equation} 
 \noindent
 The  numerical factor $\frac{\sqrt{2}\,  \psi_0}{\xi}$ becomes,
\begin{equation} 
\frac{\sqrt{2}\,  \psi_0}{\xi}= \frac{\sqrt{2}\, (1- \Omega^{1/3}_{\mathrm{m}})}{H(\tau_0)\, \tau_0} =
  \frac{(1- \Omega_{\mathrm{m}})}{\sqrt{2}}\, ,
  \end{equation}
  \noindent
  where we have used  Eq.(\ref{hx}) for $\xi$.
For $\Omega_{\mathrm{m}}=$ 0, 0.1,  0.2, 0.3, 0.5, one has $\frac{\sqrt{2}\,  \psi_0}{\xi}= $ 0.707, 0.636, 0.566, 0.495, 0.354.
Thus,  the weak field 
 equation (\ref{Nvec4})   leads to  relation (\ref{g1})  equivalent to the deep-MOND Limit 
 $g \, = \, \sqrt{a_0 \; g_{\mathrm{N}}}$.
 In this second approach,  we  may  learn much  more on the  significance  of $a_0$ and its numerical value.

 \subsection{The significance of the $a_0$-parameter}
 
We have the correspondence
\begin{equation}
a_0 \, \Longleftrightarrow  \, \frac{(1-\Omega_{\mathrm{m}})^2}{2}  g_{\mathrm{c}} \, .
\label{aog}
\end{equation}
We may explicit the limiting value  $  g_{\mathrm{c}}$ in term of the critical density 
over the radius $r_{\mathrm{H_0}}$  of the Hubble sphere,
defined by $ n\, c \, = r_{\mathrm{H_0}} H_0$. There, $n$ depends on the cosmological model. As an example, 
for the EdS model, $n=2$, for SIV or $\Lambda$CDM  models with $\Omega_ {\mathrm{m}}= 0.2 - 0.3$, the initial braking and 
 recent acceleration almost compensate each other, so that  $n \simeq 1$. We get,
 \begin{eqnarray} 
 a_0 \, = \, \frac{(1-\Omega_{\mathrm{m}})^2}{2} \, \frac{4  \pi}{3} G  \varrho_{\mathrm{c}} r_{\mathrm{H_0 }} =
 \frac{(1-\Omega_{\mathrm{m}})^2}{4} \, n\,  c \, H_0\,  \label{aog2}\\
 \mathrm{or} \quad  a_0 \,= \, \frac{n\, c \, (1-\Omega_{\mathrm{m}}) (1-\Omega^{1/3}_{\mathrm{m}})}{2\, \tau_0}
 \label {aog3}
 \end{eqnarray}
 The product $c \, H_0$ is equal to 6.80 $10^{-8}$ cm s$^{-2}$.
 For $\Omega_{\mathrm{m}}$=0, 0.10, 0.20, 0.30 and 0.50, 
   we get $a_0 \approx$ (1.70, 1.36, 1.09, 0.83, 0.43) $\; \cdot \;  10^{-8}$ cm s$^{-2}$ respectively.
 These values obtained from the SIV theory are remarkably close to the value 
   $a_0$ about 1.2 $\cdot \,10^{-8}$ cm s$^{-2}$  derived from  observations by \citet{Milgrom15}.   \\
 
 Let us make several remarks on  the $a_0$-parameter and its meaning:
 
  - 1. The  equation of the deep-MOND limit is reproduced by the SIV theory both  analytically  and numerically
  if $\lambda$ and $M$ can be considered as  constant. This may apply to systems with a typical dynamical 
  timescale  up to a few hundred million years.
 
 
 -2. Parameter $a_0$ is not a universal constant.  It depends  on the Hubble-Lema\^{i}tre $H_0$ parameter 
(or  the age  of the Universe) and on $\Omega_{\mathrm{m}}$ in the model  Universe, 
 cf. Eq. (\ref{aog2}). The value of $a_0$  applies to the present epoch. 

 -3 . Parameter $a_0$ is defined by the condition that $x >1$, {\emph{i.e.}} when the dynamical 
 gravity  $(\psi_0 \upsilon)/ \tau_0)$  in the equation of motion  (Eq. \ref{Nvec4}) becomes larger than the Newtonian gravity.
 This situation may occur over large regions at the edge of gravitational systems.


\section{Conclusions}
 The basic properties of SIV and its   relations with the scalar-tensor theories of gravity have been reviewed. Similarities and differences
 have been enlightened. 
The deep-MOND limit is found to be  an  approximation of the SIV theory for low enough densities 
and for systems with timescales  smaller than a few Myr.  

SIV theory preserves the physical properties of General Relativity and enlarges the group of symmetries 
 by the  inclusion  of scale covariance \citep{Dirac73}. 
 In fact, this also   avoid the call to dark matter, and at the same time as shown by Fig.1 the SIV
theory predicts an accelerated expansion.  Results of a number of cosmological and astrophysical applications
have been quoted in the introduction.

 Finally, while the present  approach may  for the moment  look  ``non standard''  or out of the main stream, 
 we point out that the present work  is not in contradiction with  the following statement by  \citet{Einstein49}: {\it{
 ``…the existence of rigid standard rulers is  an
assumption suggested by our approximate
experience, assumption which is arbitrary in its
principle''.}}

 \section*{Acknowledgements}
 I express my gratitude to Dr. Vesselin Gueorguiev for many years of support and excellent collaboration.\\
 
\section*{Data availability}
No new data were generated or analyzed in this research.

 \appendix
 
\section{The Hubble constant in current units}

Equation (\ref{HM}) gives the Hubble-Lema\^{i}tre parameter $H(t)$ in a timescale $t$ 
varying from the initial time $\Omega^{1/3}_{\mathrm{m}}$ to 1. We want to express it in timescale $\tau$ varying from 0 to 13.8 Gyr.
We first have  according to Eq.(\ref{dT1}),
\begin{equation}
H(\tau) \,= \, \frac{\dot{a}(\tau)}{a(\tau)}= \frac{da}{d\tau}\frac{1}{a}= \frac{da}{dt} \frac{dt}{d\tau}\frac{1}{a}=
H(t) \frac{1- \Omega^{1/3}_{\mathrm{m}}}{\tau_0}\,   \left[\frac{km}{s \cdot Mpc }\right],
\end{equation}
\noindent
For $\tau_0=$ 13.8 Gyr, the inverse $1/\tau_0$ expressed in the current units for $H_0$  
is 70.85 km/(s \, Mpc), thus we get,
\begin{equation}
H(\tau)= 70.85 \,\cdot  H(t)  (1-\Omega^{1/3}_{\mathrm{m}}) \,,
\end{equation}
\begin{equation} 
 \mathrm{with}
 \; H(t) = \frac{2 \, t^2}{t^3- \Omega_{\mathrm{m}}}\, 
 \end{equation}
 \noindent
 and relations (\ref{T2}).
  For the Hubble-Lema\^{i}tre parameter $H(\tau_0)$ at the present time  $\tau_0$, this becomes
 \begin{equation}
H(\tau_0)= 2 \,\cdot  \,70.85 \; \frac{1-\Omega^{1/3}_{\mathrm{m}}}{1-\Omega_{\mathrm{m}} }  \;\left[ \frac{km}{s \; Mpc} \right] \,.
 \end{equation}
 For $\Omega_{\mathrm{m}}=0,$  0.10, 0.20, 0.30, 0.50, the value of $\xi = (H(\tau_0) \cdot \tau_0) $ are 2, 
 1.191, 1.038, 0.945, 0.814  respectively (see Sect.  \ref{2ap}). Interestingly enough,  for 
$\Omega_{\mathrm{m}}=1,$ we have $\xi=2/3$, scale invariance disappears and as expected the flat  model converges 
towards the EdS model. The above corresponding values of $H(\tau_0)$ are 141.70, 84.38, 73.54, 66.95, 57.70 in km/(s Mpc).

\bsp	
\label{lastpage}
\end{document}